\documentclass[prx,english,superscriptaddress,longbibliograph,twocolumn]{revtex4-2}
\usepackage{graphicx} 
\usepackage{amsmath}
\usepackage{braket}
\usepackage{bm}
\usepackage{tabularx}
\usepackage{array}
\usepackage{ulem}
\newcolumntype{P}[1]{>{\centering\arraybackslash}p{#1}}

\newcolumntype{Y}{>{\centering\arraybackslash}X}

\DeclareMathOperator{\tr}{tr}
\DeclareMathOperator{\mathspan}{span}

\begin{document}
\title{The Floquet Fluxonium Molecule: \\ Driving Down Dephasing in Coupled Superconducting Qubits}

\author{Matthew Thibodeau}
\email{mt24@illinois.edu}

\affiliation{Department of Physics, University of Illinois Urbana-Champaign, Urbana, IL, United States 61801}
\affiliation{IQUIST and Institute for Condensed Matter Theory and NCSA Center for Artificial Intelligence Innovation, University of Illinois at Urbana-Champaign, IL 61801, USA}
\author{Angela Kou}

\affiliation{Department of Physics, University of Illinois Urbana-Champaign, Urbana, IL, United States 61801}
\affiliation{Materials Research Laboratory,
University of Illinois at Urbana-Champaign, Urbana, IL 61801, USA}
\author{Bryan K. Clark}

\affiliation{Department of Physics, University of Illinois Urbana-Champaign, Urbana, IL, United States 61801}
\affiliation{IQUIST and Institute for Condensed Matter Theory and NCSA Center for Artificial Intelligence Innovation, University of Illinois at Urbana-Champaign, IL 61801, USA}

\date{\today}
\begin{abstract}
    High-coherence qubits, which can store and manipulate quantum states for long times with low error rates, are necessary building blocks for quantum computers. Here we propose a driven superconducting erasure qubit, the Floquet fluxonium molecule, which minimizes bit-flip rates through disjoint support of its qubit states and suppresses phase-flips by a novel second-order insensitivity to flux-noise dephasing.
    We estimate the bit-flip, phase-flip, and erasure rates through numerical simulations, with predicted coherence times of approximately 50~ms in the computational subspace and erasure lifetimes of about 500~$\mu$s. We also present a protocol for performing high-fidelity single-qubit rotation gates via additional flux modulation, on timescales of roughly 500~ns, and propose a scheme for erasure detection and logical readout. Our results demonstrate the utility of drives for building new qubits that can outperform their static counterparts.
\end{abstract}

\maketitle

\section{Introduction}

High coherence qubits with good protection from environmental noise are a key enabling technology for fault-tolerant quantum computers. Such qubits can perform more operations before an expected error in the encoded quantum state and thus will require less overhead from quantum error correction (QEC).  Protected qubits with long coherence times for Pauli-type errors 
have been of significant interest, but 
have so far been difficult to implement due to the need for extreme parameter regimes. 

Pauli-type errors are relatively difficult to handle via QEC. By contrast, erasure-type errors are much more benign, as quantum codes exist with much higher fault tolerant thresholds for erasures \cite{Grassl1997, Kubica2023, Wu2022}. 
This motivates the design of protected erasure qubits, where Pauli errors occur at very low rates with moderate levels of erasures. The key challenge is to engineer an interacting system that gives the desired erasure-biased error hierarchy. 

This engineering can be done using circuit QED where Hamiltonians can be quite flexible \cite{Blais2021}, although certain species of neutral atoms are also natural erasure qubits \cite{Wu2022}. Circuit QED in particular allows multiple-degree-of-freedom (DOF) Hamiltonians which can have very favorable coherence properties. This category includes protected qubits such as the $0-\pi$ and cold echo qubits \cite{Brooks2013,Gyenis2021zeropi,Groszkowski2018,Kapit2022}, as well as erasure qubits including dual-rail cavities \cite{Bajjani2011,Teoh2023,Chou2024} and the dual-rail transmon \cite{Kubica2023}. Early experiments on the latter have achieved relatively long logical coherence times with strong erasure error biases of up to $30\times$ \cite{Levine2023}. However, erasure qubits demonstrated thus far have exhibited strictly worse overall coherence than the best single-DOF qubits \cite{Somoroff2023}.

Here we propose a novel protected erasure qubit, the Floquet fluxonium molecule (FFM). The FFM qubit exhibits (i) extremely long predicted logical coherence times and relatively long erasure lifetimes, (ii) a simple superconducting circuit structure, and (iii) high-fidelity single qubit gates which are much faster than the coherence timescale. Based on a Floquet-driven pair of inductively coupled fluxonium circuits \cite{Manucharyan2009, Kou2017, Somoroff2023}, the FFM is a multi-DOF superconducting circuit with engineered, highly coherent quasi-eigenstates.

Our key technical contribution is a novel form of Floquet protection in a multi-DOF qubit which strongly suppresses phase-flip errors, removing them at first- and second-order in the flux noise. The combination of drive and multi-DOF allows the low-lying eigenstates to be disjoint and delocalized with a non-vanishing energy gap. The second-order sweet spot has no analogue in the single-DOF circuits that have been studied thus far \cite{Mundada2020,Didier2019,Hong2020}; in fact, in single-DOF circuits there is a generic trade-off between bit- and phase-flip errors arising from the inability to keep two eigenstates simultaneously disjoint and flux-delocalized using accessible circuit QED Hamiltonians \cite{Gyenis2021}. 

The higher-order phase-flip insensitivity allow the predicted coherence time of the FFM qubit to significantly outperform other multi-DOF circuits. These include: the dual-rail erasure transmon, with experimentally achieved logical lifetimes of $\approx 1$~ms and erasure lifetimes of $\approx 30$~$\mu$s \cite{Levine2023}; the dual-rail cavity, with logical lifetimes predicted \cite{Teoh2023} (achieved \cite{Chou2024}) at $\approx 10$~ms ($3$~ms), limited by cavity and ancilla dephasing, and erasure lifetimes of $\approx 500$~$\mu$s in both cases; and the cold echo qubit, with predicted logical lifetime of $T_L \approx 16$~ms with erasure rates unreported \cite{Kapit2022}.
Theoretically, we find the FFM exhibits long bit-flip coherence times of approximately 50~ms while suppressing phase-flips even further, along with a 500~$\mu$s erasure lifetime. The Floquet-induced protection of the FFM logical subspace does not require fine-tuning of the static Hamiltonian, which allows additional flexibility in fabrication and parameter selection. Furthermore, the protection scheme is general enough for potential application to other types of qubit hardware.

\subsection{Design motivation}
In this work, we develop a driven qubit based on the fluxonium circuit that exhibits dephasing suppressed to second order and, simultaneously, suppressed bit-flips due to its flux-disjoint computational states.
To accomplish both goals together, more than one physical degree of freedom is required \cite{Gyenis2021}.
Directing our attention to multi-DOF circuits, we in particular focus on the circuit shown in Figure \ref{fig:introfigure} \cite{Kou2017,Kapit2022}, known as the fluxonium molecule. The Hamiltonian is defined in equation (\ref{eq:FFMHam}) with external fluxes $\phi_{L} = \phi_{R} = \pi$, and the dynamical degrees of freedom are $\varphi_L$ and $\varphi_R$.
This circuit nearly satisfies both coherence goals if the $g$ and $f$ states are taken as the computational basis states. They have disjoint wavefunctions to minimize bit-flips, and if we expand the shift in the qubit frequency $\epsilon_{01}$ due to some flux noise $\delta \phi_j$ using perturbation theory
\begin{align}\label{eq:o_op}
    \delta \epsilon_{10} = \delta\phi_j \Delta^{(1)}_{j} + \left(\delta\phi_j\right)^2 \Delta^{(2)}_{j} + \dotsm
\end{align}
then this system satisfies $\Delta^{(1)} = 0$ for both of its flux-mode dephasing error channels $\delta\phi_L$ and $\delta\phi_R$; we say that it is protected from dephasing up to first order in the noise. However, there is a problem: $\Delta^{(2)}$ is very large for this undriven fluxonium molecule, due to vanishing energy gaps in the denominator (see equation (\ref{eq:secondordergeneral})). To make the $g$ and $f$ states nearly disjoint, the tunneling energy $E_C$ must be low, which necessarily makes the $g-e$ subspace and the $h-f$ subspace nearly degenerate.  
Similarly, considering the qubit frequency dispersion of this circuit with respect to the flux noise channels $\delta\phi_L$ and $\delta\phi_R$, one finds that it is quadratic over a very small parameter range around zero noise and linear just beyond that; such a dispersion is very sensitive to becoming linear at a finite nonzero noise fluctuation amplitude, which gives rise to the aforementioned very large second order term when Taylor-expanded around zero. 

In principle, an interaction diagonal in the four low-lying states, $g,e,f,h$, could be used to directly tune the energy splittings of the states and eliminate the vanishing denominators. We are not aware of any simple circuit element which could provide such an interaction; we therefore use Floquet engineering to properly adjust the four low-lying eigenstates. 

Floquet engineering relies on modulating an operator $\mathcal O$ in the Hamiltonian at frequency $\Omega$ to effectively introduce an interaction between eigenstates of $H$, resulting in ``copies'' of those eigenstates with their energies shifted by $\pm \Omega$ (see Appendix \ref{floquetlattice}) \cite{Oka2019,Long2022,Mundada2020}. Resonances, which are tunable through their dependence on $\Omega$, can generate interactions that would be difficult or impossible to produce using non-driven electrical components. In the rest of this paper, we will show that with Floquet engineering it is possible to dramatically increase the phase-flip coherence of the fluxonium molecule circuit while preserving its high bit-flip coherence.

\section{Diagonalizing the FFM model}
\begin{figure}
    \centering
    \includegraphics[width=\linewidth]{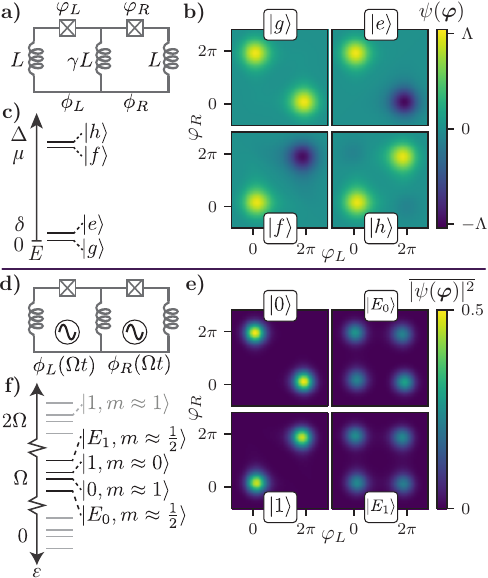}
    \caption{
    (a) The static fluxonium molecule circuit. We use the Hamiltonian parameters from Table \ref{tab:circuitparam_table} with left and right external fluxes set to $\phi_L = \phi_R = \pi$. The wavefunctions $\psi_j(\bm{\varphi})$ of the four lowest eigenstates $\ket{j}$ are shown in (b); the colorbar scale is $\Lambda = \max \psi \approx 0.672$. There are two pairs of eigenstates with mutually disjoint wavefunctions, and (c) shows a sketch of their energy levels.
    Floquet physics is introduced to the fluxonium molecule in (d) by driving the external flux monochromatically at amplitude $A$ and frequency $\Omega$; see eq. (\ref{eq:driveeqn}). (e) The corresponding wavefunctions, now plotting the time-averaged value of $|\psi(\bm\varphi,t)|^2$, showing how the four low-lying states are mixed by the drive into two disjoint computational states and two erasure states. (f) The quasi-energy sketch for the driven Floquet system annotated with their average photon number $m$. In light gray are the nearest two copies of the quasi-eigenstates in the frequency lattice picture (see Appendix \ref{floquetlattice}).
    }
    \label{fig:introfigure}
\end{figure}
We consider the fluxonium molecule (FM) superconducting circuit \cite{Kou2017} shown in Figure \ref{fig:introfigure}(a), which supports two dynamical variables $\varphi_L, \varphi_R$ and their canonical conjugates $n_L ,n_R$. An external, time-dependent classical flux $\Phi^{ext}_{L,R} = \tfrac{\Phi_0}{2\pi} \phi_{L,R}(t) $ threads each loop of the circuit, and we thus define the offset variables $\overline \varphi_j(t) = \varphi_j - \phi_j(t)$. The (time-dependent) Hamiltonian is \cite{You2019}

\begin{equation}
\begin{split}\label{eq:FFMHam}
    H_{FFM}(t) &= 4E_C\left(n_L^2 + n_R^2\right) + \frac 1 2 E_L \left(\overline \varphi_L^2 + \overline\varphi_R^2\right) \\ &+ \frac 1 2 E_L' \overline \varphi_L \overline \varphi_R - E_J(\cos\varphi_L + \cos\varphi_R)
\end{split}
\end{equation}
where $E_C$ and $E_J$ are the charging Josephson energies, respectively, and $E_L$ and $E_L'$ are inductive energies satisfying
\begin{align}
    E_L = \frac{1 + \gamma}{L(1 + 2\gamma)} \quad \quad E_L' = \frac{2\gamma}{L(1 + 2\gamma)}
\end{align}
with $\gamma$ as shown in Figure \ref{fig:introfigure}(a).
This model has been experimentally realized in its static form and has also been analyzed in the presence of time-dependent charge driving fields \cite{Kapit2022}. For our purposes we will require that $E_L, E_C \ll E_J$. Below we will describe a new method of engineering highly coherent Floquet eigenstates by resonantly driving the external fluxes $\phi_{L,R}$.

We define the common-mode and differential-mode flux operators by $\varphi_{C} = \tfrac{1}{2}(\varphi_L + \varphi_R)$ and $\varphi_D = \varphi_R - \varphi_L$, as well as their associated offset fluxes $\phi_{C,D}(t)$. We fix a static common-mode flux, operating the individual fluxoniums at their half-flux regime, so the time-dependence of $H_{FFM}$ will enter through a drive of the differential flux $\phi_D(t)$. In particular, we work with a monochromatic drive of amplitude $A$:
\begin{align}\label{eq:driveeqn}
    \phi_D = 2\pi A\sin\Omega t \quad\quad \phi_C = \pi
\end{align}
For a specific choice of $A^*, \Omega^*$, we will have two nearly disjoint eigenstates with $\Delta^{(1)}=\Delta^{(2)}=0$ for flux noise. These two eigenstates can then be used as a qubit.

\begin{table}
        
    \begin{tabularx}{0.45\textwidth}{Y*{5}{|Y}}
 & $E_C$ & $E_J$ & $E_L$ & $E'_L$ & $g$\\ \hline 
         (GHz)  & 0.7 & 3.9 & 0.4 & $0.20667$ & 0.25 \\ \hline
    \end{tabularx}
    \caption{Circuit parameters for the numerically diagonalized $H_{FFM}$.}
    \label{tab:circuitparam_table}

    \vspace{0.5cm}
    \begin{tabular}{p{2.5cm} | P{2cm} | P{2cm}| P{1.25cm}}
 \hfill($\mu$s) & $T_1$ & $T_{\phi, \varphi}$  & $T_e$ \\ \hline 
         FFM Qubit& $49.0 \times 10^3$ & $227 \times 10^3$  & 524 \\ \hline
         Static FM Qubit & $38.2 \times 10^3$ & 682 & 281 \\ \hline
    \end{tabular}
    \caption{Coherence times for the FFM qubit, compared to the static fluxonium molecule (FM) with $\ket{0} = \ket{g}$ and $\ket{1} = \ket{f}$. The drive dramatically improves the dephasing coherence time $T_{\phi,\varphi}$ and slightly increases the erasure coherence time $T_e$ while retaining a high depolarization time $T_1$.
    }
    \label{tab:coherence_table}

\end{table}

\subsection{Static low-energy spectrum}
Consider the static, half-flux scenario when $\phi_D = 0$ and $\phi_C = \pi$. Here, the FM Hamiltonian, when $E_L \ll E_J$, is analogous to that of a particle moving in the 2D plane $\bm{\varphi} = (\varphi_L, \varphi_R)$ subject to a four-well potential, with minima near $\varphi_L,\varphi_R \in \{0, 2\pi\}$. 
The interacting $\varphi_L\varphi_R$ term in $H_{FFM}$ splits the two diagonal wells with $(\varphi_L, \varphi_R) \approx (0,0)$ or $(2\pi, 2\pi)$ from the remaining antidiagonal wells by an energy gap proportional to $E'_L$. In the regime of low tunneling, $E_C \ll E_J$, the eigenstates have $\varphi$ wavefunctions approximately localized within these four wells as shown in Figure \ref{fig:introfigure}(b). The eigenstate $\ket{g}$ (resp. $\ket{e}$) is approximately a symmetric (antisymmetric) superposition over the antidiagonal wells; similarly, the next two eigenstates $\ket{f},\ket{h}$ are $\pm$ superpositions over the diagonal wells.
Each doublet has an energy splitting due to the kinetic term $E_C$.
Additionally, at nonzero tunneling ($E_C > 0$), the eigenstates weakly mix between the diagonal and antidiagonal wells, which will become important in the next section when we diagonalize $H_{FFM}$ perturbatively in the mixing strength.

The resulting low-energy spectrum of the FM is shown in Figure \ref{fig:introfigure}(c), with energies $0, \delta, \mu, \Delta$ for the $g,e,f,h$ states. From the above considerations we expect a hierarchy $\delta, \Delta -\mu \ll \mu, \Delta$ and that the $g,e$ wavefunctions have approximately disjoint support from the $f,h$ wavefunctions. Calculating the energies directly given $H_{FM}$ is not generally possible analytically, and so we rely on exact diagonalization of a truncated $H_{FM}$ to obtain their quantitative values given the circuit parameters $E_C, E_J, E_L, E_L'$.

Higher levels exist, but are separated anharmonically from the low-energy states and thus do not strongly participate in any resonant phenomena among those states; we investigate their effects in section \ref{manylevelED}. As a first approximation we truncate the FM Hilbert space to $\mathcal H_4 = \mathspan \{g,e,f,h\}$, but we include the effects of higher levels later in our numerical analysis. 

\begin{figure}
    \centering
    \includegraphics[width=\linewidth]{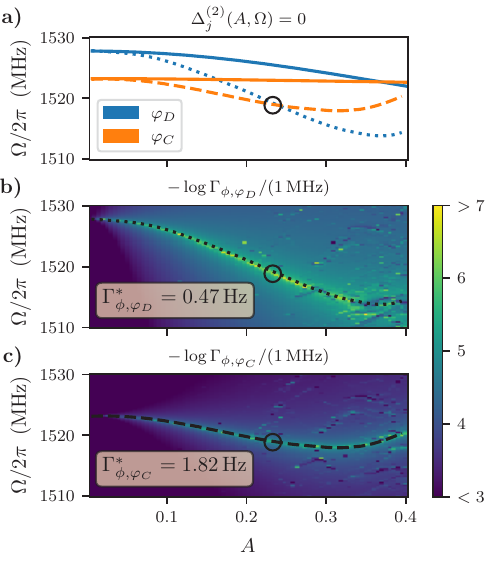}
    \caption{(a) Curves in the $A,\Omega$ drive parameter plane where each of the second-order contributions to dephasing $\Delta_{j}^{(2)}$, from differential ($j = D$) and common ($j = C$) flux noise, are zero in the 4-level approximation (solid lines) and converged many-level ($N = 300$) approximation (dashed lines).
    The encoded qubit is maximally insensitive to dephasing from flux noise when $\Delta^{(2)}_j = 0$ for both $j = C,D$. For each approximation, there is one point $(A^*,\Omega^*)$ where the curves cross, and here the qubit is second-order insensitive to dephasing from both flux noise sources. (b,c) Dephasing coherence plots of the FFM qubit for flux noise-induced errors, from sources (b) $\phi_D$ and (c) $\phi_C$. The high-coherence areas closely track the many-level $\Delta^{(2)} = 0$ lines from (a), which are overlaid. The marked point indicates the predicted  maximal dephasing-coherence point, and the calculated dephasing rates (within the resolution of the grid) at that point are shown in the lower left. 
    } 
    \label{fig:perturbationtheory}
\end{figure}

\subsection{Floquet eigenstates in the 4-level theory}\label{4leveltheory}
We generalize to the case where the differential flux drive amplitude is nonzero, $A > 0$, and discuss the Floquet quasi-eigenstates in the aforementioned 4-level approximation. Focusing on these four states allows us to analytically obtain the Floquet quasi-eigenstates and quasi-energies by doing perturbation theory in a secondary parameter $\epsilon \propto \braket{e|\varphi_D|h}$, as defined in equation (\ref{eq:phidbprimebasis}). This parameter $\epsilon$ corresponds to the mixing between states living in the antidiagonal and diagonal wells that occurs at nonzero tunneling energy $E_C$. 

The matrix of $\varphi_D$ is otherwise sparse in the low-energy basis, with zero diagonal and zero coupling to the state $\ket{f}$. This structure is not fine-tuned, and is a consequence of the simple structure of the static, non-driven 4-level theory: effectively, four deep wells in a square lattice with a potential energy difference between the two diagonals, weak tunneling between nearest neighbors, and much weaker tunneling across diagonals. The Hamiltonian with this much-simplified spatial structure reproduces qualitatively the eigenstates and energies of Figure \ref{fig:introfigure}(b-c) as well as the matrix element structure of $\varphi_C$ and $\varphi_D$. When going beyond the 4-level theory, additional coupling can be introduced with higher energy static eigenstates. We will address this numerically in section \ref{manylevelED}, finding the same qualitative phenomena with slightly different quantitative predictions.

We choose to drive the system at frequency $\Omega \approx \Delta - \delta$. This choice effectively brings the $\ket{e}$ and $\ket{h}$ states into resonance, and causes them to hybridize; the degree to which they hybridize is controlled by $A$ and $\Omega$. We give the full details of the analysis in Appendix \ref{diagonalizeK}, but here we highlight some important qualitative features of the Floquet eigenstates.

The four static eigenstates are mixed by the drive into four time-dependent quasi-eigenstates which we separate into two subspaces: the computational subspace spanned by $\ket{0}$ and $\ket{1}$, and the erasure subspace spanned by $\ket{E_0},\ket{E_1}$. We define $\varphi$-wavefunctions for each state $\ket{\alpha(t)}$ in the usual way:
\begin{align}
    \psi_{\alpha}(\bm\varphi, t) = \braket{\bm{\varphi}|\alpha(t)}
\end{align}

The driven states $\ket{0}$ and  $\ket{1}$ are closely connected to the undriven states $\ket{g},\ket{e}$ and $\ket{f}$; in this 4-level approximation, we may summarize their relationship by
\begin{align}
    \ket{1} &= e^{i\Omega t}\ket{f} \\
    \ket{0} &\approx \frac{1}{4}\left((\ket{g} + \ket{e}) + e^{i\theta(t)}(\ket{g} - \ket{e})\right)
\end{align}
where the phase function $\theta(t)$ is periodic with frequency $\Omega$ and has a frequency expansion given by a sum of Bessel functions (see equations (\ref{eq:lowamp_xyzwstates_x}) and (\ref{eq:lowamp_zerostate}) for details). This implies that the support of the computational state wavefunctions very nearly correspond to those of the $g/e$ and $f$ states, with $|\psi_0(t)|^2 \approx |\psi_{g}|^2 \approx |\psi_{e}|^2$ and $|\psi_1(t)|^2 \approx |\psi_f|^2$ at all times $t$, as is suggested by Figures \ref{fig:introfigure}(b) and \ref{fig:introfigure}(e). In particular, $\psi_0$ and $\psi_1$ have nearly disjoint support. The Floquet eigenstate wavefunctions $\psi_0, \psi_1$ do carry additional time- and $\bm \varphi$-dependent phases relative to the static states. 

Close to the resonance condition $\Omega \approx \Delta - \delta$, the erasure states $\ket{E_0},\ket{E_1}$ are time-dependent but near-uniform superpositions of the remaining state $h$ with the remaining orthogonal $g/e$ superposition state, again carrying time- and $\bm \varphi$-dependent phases relative to the static states. The quasi-energies $\epsilon_{E0}$ and $\epsilon_{E1}$ depend on the drive and circuit parameters, but close to the resonance condition they are structured above and below the computational quasi-energies $\epsilon_{0,1}$ as shown in Figure \ref{fig:introfigure}(f). The qubit frequency is $\epsilon_{10} = \epsilon_1 - \epsilon_0$.

Now we can address dephasing by evaluating the effect of noisy operator fluctuations on $\epsilon_{10}$ at second order:
\begin{align}\label{eq:secondordergeneral}
    \Delta^{(2)}_{j} = \sum_{\alpha \neq 1}\frac{|\varphi_j|^2_{1\alpha}}{\epsilon_1 - \epsilon_\alpha} - \sum_{\alpha \neq 0}\frac{|\varphi_j|^2_{0\alpha}}{\epsilon_0 - \epsilon_\alpha}
\end{align}
The sum in equation (\ref{eq:secondordergeneral}) runs over the Floquet quasi-eigenstates. Given these and their quasi-energies, we calculate the second-order noise energy shifts $\Delta^{(2)}_{C,D}(A,\Omega)$ as a function of the drive parameters, and then solve the two equations $\Delta^{(2)}_{C,D} = 0$ in Appendix \ref{lowampappendix}. The result, shown in solid lines in Figure \ref{fig:perturbationtheory}(a), are two curves in the $(A,\Omega)$ plane, one each for $\varphi_C$ and $\varphi_D$, where external flux fluctuations coupling these operators have minimal effect on the qubit energy. An illustration of the energy dispersion with respect to flux is shown in Figure \ref{fig:dispersion}, showing that the FFM eigenstates with $\Delta^{(2)}_{C,D} \approx 0$ have essentially no dispersion when compared to the static FM eigenstates possessing large second-order responses.

As we will see in the next section, on the curves in Figure \ref{fig:perturbationtheory}(a) the dephasing rate of the qubit due to flux fluctuations is very low, as expected.
For our circuit parameters, the curves cross at finite and nonzero $A$, implying the existence of a protected ``double sweet spot'' of drive parameters $(A^*, \Omega^*)$ where the qubit frequency is maximally and simultaneously insensitive to both the $\phi_C$ and $\phi_D$ flux noise channels.

\subsection{Many-level exact diagonalization}\label{manylevelED}
The 4-level theory furnishes a simple prediction for the curves of $\Delta^{(2)}_j = 0$. However, in reality the (static) Hamiltonian $H_{FM}$ supports infinitely many eigenstates. Two effects necessitate the inclusion of higher levels in a quantitative analysis of qubit coherence: weak mixing of low-lying excited states, which occurs even at low $A$, and accidental multi-photon resonances that occur at larger $A$. 

Low-lying excited states are off-resonant with respect to the chosen drive, but do still weakly mix with the $g,e,f,h$ states at any nonzero drive amplitude. This effect slightly modifies the computational and erasure quasi-energies and quasi-eigenstate matrix elements obtained from the 4-level approximation, and hence shifts the $\Delta^{(2)} = 0$ solution curves. To accurately predict the double sweet spot parameters $(A^*, \Omega^*)$, the shift must be computed numerically.

A second detrimental effect arises due to multi-photon resonances with highly excited states, which becomes an issue at larger drive amplitude. These resonances, visible in the coherence data of Figure \ref{fig:perturbationtheory}(b-c) above $A \approx 0.3$, can potentially cause strong hybridization of the computational and erasure states, destroying the cancellation of the $\Delta^{(2)}_j$. For instance, suppose a high-lying excited state $\ket{\alpha}$ has energy $E_\alpha = m \Omega - \eta_\alpha$ for some integer $m$ and small $\eta_\alpha$. Then it is nearly on-resonance for an $m$-photon Floquet transition in the frequency lattice picture (see Appendix \ref{floquetlattice}) from the ground state $g$ with energy $E_g = 0$. If $\eta_\alpha = 0$ and the degeneracy is exact, hybridization will occur and the $\Delta^{(2)}$ coefficients will differ from the 4-level theory prediction.

Both effects --- weak mixing of low-lying states and accidental $m$-photon degeneracies --- are quantified by exactly diagonalizing the Floquet Hamiltonian. We truncate $H_{FFM}(t)$ in the frequency lattice picture up to a Fourier cutoff $M$ and a static cutoff $N$ and numerically find its quasi-eigenstates and quasi-energies for a range of $\Omega$ and $A$. Given this data we then solve for the zero locus of $\Delta^{(2)}_j$ via eq. (\ref{eq:secondordergeneral}). Results for our circuit parameters are in Figure \ref{fig:perturbationtheory}(a), where we show in dashed lines the numerically computed $\Delta^{(2)}_j = 0$ curves alongside the 4-level theory prediction in solid lines. We provide evidence for convergence of our simulations in Appendix \ref{convergence}.

Although the $\Delta^{(2)}_j = 0$ curves are shifted, they still cross, so that the existence of a double sweet spot is preserved in the more accurate many-level picture. Additionally, we see empirically that this crossing occurs at low drive amplitude, before the widespread proliferation of strong accidental resonances. These findings suggest that the Hamiltonian supports a protected qubit in the computational subspace at the double sweet spot.

Note that this is a sweet spot of the \textit{drive} parameters, not the static circuit parameters, and that no fine-tuning of the circuit parameters is necessary to produce a double sweet spot. To this point, Figure \ref{fig:paramsweep} illustrates how the drive parameters that result in the double sweet spot change as the circuit parameters $E_J$ and $E_L'$ are tuned from the values we consider here.

\section{Coherence of the FFM qubit}\label{coherence}

\begin{figure}
    \centering
    \includegraphics[width=\linewidth]{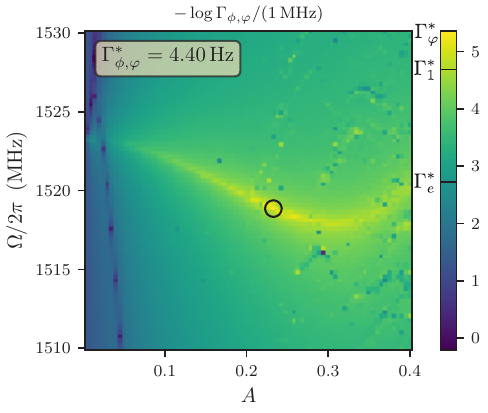}
    \caption{The total flux-dephasing (see eq. (\ref{eq:totaldephasing})) error rates computed for the FFM qubit with parameters given in Table \ref{tab:circuitparam_table}. The center of the circle $\circ$ marks the maximal dephasing coherence point with the assumed noise amplitudes. As a comparison, on the colorbar we indicate the erasure error rate $\Gamma_e^*$ and depolarization rate $\Gamma_1^*$ at the marked point.
    The individual flux noise contributions to dephasing are shown in Figure \ref{fig:perturbationtheory}.
    }. 
    \label{fig:coherence-plot}
\end{figure}

By design, the FFM qubit is protected from both dephasing and bit-flips at its operating point $(A^*, \Omega^*)$. Bit-flip insensitivity is achieved because the computational states wavefunctions $\psi_0(\bm \varphi)$ and $\psi_1(\bm \varphi)$ have disjoint support, so operators local in the flux (including both $\varphi_j$ and $n_j = -i \tfrac{\partial}{\partial \varphi_j}$) have small matrix elements connecting them. Simultaneously, phase-flip insensitivity is achieved through both first- and second-order cancellation of energy shifts due to flux noise. Overall, erasure errors are predicted to dominate the system, but still occur at relatively low rates of order kHz. These results, along with quantitative error rate estimates, are summarized in Figure \ref{fig:coherence-plot}.

\subsection{Dephasing errors}\label{dephasingsection}
In sections \ref{4leveltheory} and \ref{manylevelED} we computed the perturbative effect of flux noise on the qubit frequency and found a drive point where the noise-induced frequency shift vanishes at both first and second order; in the vicinity of this point one expects a very low dephasing rate. We verify this by estimating the qubit frequency shifts from both of the noisy external fluxes $\Phi_j$ ($j = C$ or $D$) using a finite difference calculation, subtracting the qubit frequency $\epsilon'_{10}$ after a noise excursion of characteristic size $A_{\Phi_j}$ from the expected zero-noise $\epsilon_{10}$ to obtain realistic values of the dispersion $|\delta \epsilon_{10}|_j$. We use this value as the proxy for the dephasing rate. For flux noise following a $1/|f|$ spectrum, we have
\begin{align}
    \Gamma_{\phi, \varphi_j} = |\delta\epsilon_{10}|_j \sqrt{2\log^2 \tfrac{\omega_{uv}}{\omega_{ir}} + 4 \log^2 \omega_{ir}t}
\end{align}
which includes an additional multiplicative factor to account for the logarithmic divergence of the noise spectra, with frequency cutoffs $\omega_{uv} = \Omega \approx 2\pi \times 1.5$ GHz and $\omega_{ir} = 2\pi \times 1$ Hz and integration time $t = 10\mu$s \cite{Groszkowski2018}. This approximation is valid in our regime where $\Delta^{(1)}_{C,D} = 0$. 
We similarly compute dephasing from drive amplitude noise, although we drop the logarithmic factor and use $\Gamma_{\phi,ac} = |\delta \epsilon_{01}|_{ac}$.

The total pure-dephasing decoherence rate we compute is
\begin{align}\label{eq:totaldephasing}
    \Gamma_{\phi,\varphi} = \Gamma_{\phi,\varphi_D} + \Gamma_{\phi,\varphi_C} + \Gamma_{\phi,ac}
\end{align}
where on the right hand side we have the pure-dephasing rates due to $\phi_D$-flux fluctuations, $\phi_C$-flux fluctuations, and drive amplitude fluctuations respectively. 

In Figure \ref{fig:coherence-plot}(a), we show the computed total rate $\Gamma_{\phi,\varphi}$ over a range of drive parameters for assumed flux noise strengths $A_{\Phi_C} = A_{\Phi_D} = 10^{-6}~\Phi_0$ and drive amplitude noise strength $A_{ac} = 10^{-8}$. As expected from perturbation theory, the flux noise dephasing rate is extremely low near the double sweet spot.
In Figure \ref{fig:perturbationtheory}(b) and (c) we plot the two flux noise components $\Gamma_{\phi,\varphi_D}$ and $\Gamma_{\phi,\varphi_C}$ individually. We show $\Gamma_{\phi,ac}$ in  Figure \ref{fig:amplitudedephasing}.

\subsubsection{Dephasing from flux noise}

We have seen in section \ref{4leveltheory} that there is a point in drive parameter space $(A^*, \Omega^*)$ where $\Delta^{(2)}_j = 0$ for both $j = C$ and $D$. In practice, this is a measure-zero point which is difficult or impossible to fix exactly in an experiment. Instead we use a finite-resolution grid of $dA = 0.005$ and $d\Omega = 0.25$ MHz and report results at the point closest to the true  $(A^*, \Omega^*)$, which is marked on the plots with a small circle $\circ$. For example, at this point $|\Delta^{(2)}_D| \approx 2$~GHz; this is compared to approximately $|\Delta^{(2)}_D| \approx 1.9 \times 10^4$~GHz at $A = 0$, which accounts for the increased dephasing performance of our proposed design. At this approximate double sweet spot the flux noise dephasing rate becomes very low, approximately 2.3 Hz with our circuit parameters and noise model.

\subsubsection{Dephasing from drive amplitude noise}

\begin{figure}
    \centering
    \includegraphics{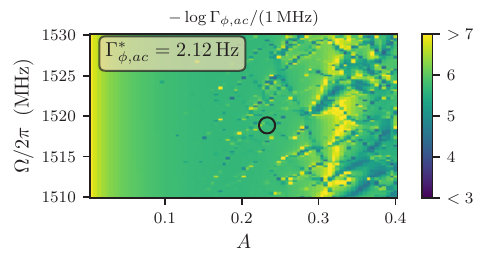}
    \caption{Dephasing rate $\Gamma_{\phi,ac}$ due to amplitude noise, along with the double sweet spot marked with $\circ$ and the amplitude dephasing rate $\Gamma_{\phi, ac}^*$ at that point. Although the qubit is linearly sensitive to amplitude fluctuations, those fluctuations are of very low amplitude in a good-quality signal generator, so the net dephasing impact is expected to be comparable to that from flux noise.}
    \label{fig:amplitudedephasing}
\end{figure}

The Floquet qubit introduces two new parameters, the drive amplitude $A$ and frequency $\Omega$, which are in general noisy sources of dephasing. Here we will assume that the $\Omega$ fluctuations are small enough to ignore, and instead focus on drive amplitude noise. 
The computed dephasing from drive amplitude noise is shown in Figure \ref{fig:amplitudedephasing}.

In the 4-level picture, we can calculate the qubit frequency dispersion with respect to $A$. The result, obtained in Appendix \ref{lowampappendix}, is:
\begin{align}\label{eq:amplitude_dispersion}
    \frac{\partial \epsilon_{01}}{\partial A} \propto \epsilon^2 \left( f_+'(A/A_0) + f_-'(A/A_0)\right) + (r-2\epsilon^2)J_{1}(2A/A_0)
\end{align}
where the drive normalization $A_0$ is defined in (\ref{eq:a0eq}), $r = \delta/\Delta$, and the functions $f_+$ and $f_-$ are defined in equation (\ref{eq:fplusminus}). 
This dispersion has an overall scaling with $\epsilon^2$ and $r$; these depend only on the static circuit parameters, so that choosing parameters with small $r$ and $\epsilon$ will increase the amplitude noise coherence time. 

On the other hand, for any given $r$ and $\epsilon$ one can set eq. (\ref{eq:amplitude_dispersion}) equal to zero and search for a solution at which the drive strength $A'$ has only second-order dephasing; carefully tuning the circuit parameters can allow one to fix $A' = A^*$ so that the first-order protected point for drive amplitude noise coincides with the second-order point for flux noise. 

In practice, we expect the amplitude noise dephasing rate to be on par with that from flux noise even in the linear-dispersive regime where $\frac{\partial \epsilon_{01}}{\partial A} \neq 0$, and therefore we did not tune our circuit parameters to achieve this matching condition.
The drive amplitude noise is expected to be very low from good-quality signal generators, with power spectral density below -120 dBc/Hz when using a GHz-range carrier $\Omega$ as is the case here. Moreover, the dephasing rate due to the drive amplitude is expected to be subdominant relative to the erasure rate in the linear-dispersive regime.

In addition, drive amplitude (and drive frequency) fluctuations are, at least in principle, amenable to active correction. Unlike $\phi_C$ and $\phi_D$ noise, which arise from essentially unobservable fluctuations within a device, the drive signal is generated by the experiment and can be classically monitored to arbitrarily high sensitivity, limited ultimately by thermal noise in the drive lines. This raises the possibility of actively correcting phase errors induced by amplitude or frequency fluctuations, either by adjusting the drive through feedback or by actively updating the frequency of the qubit's rotating frame. 

\subsection{Depolarization from flux and charge noise}
\begin{figure}
    \centering
    \includegraphics{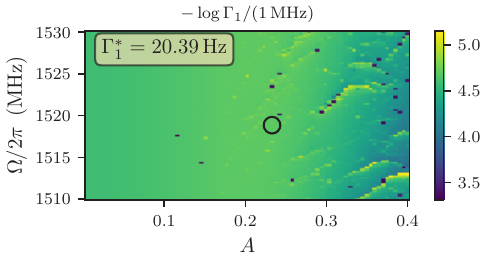}
    \caption{The depolarization $\Gamma_1$ error rate computed for the FFM qubit, with the double sweet spot marked with $\circ$ and the depolarization rate $\Gamma_{1}^*$ at that point. The erasure error rate is roughly constant at $\Gamma_{e} \approx 1.91$~kHz.}
    \label{fig:depolcoherence}
\end{figure}

The computed bit-flip rate due to flux and charge noise is shown in Figure \ref{fig:depolcoherence}. For a static qubit, the decoherence rate associated to an operator $\mathcal O$ is modeled by Fermi's golden rule
\begin{align}
    \widetilde\Gamma_{1\mathcal O} = \frac{1}{\hbar^2}\left|\braket{0|\mathcal O | 1}\right|^2 S_{\mathcal O}(\epsilon_{01})
\end{align}
where $S_{\mathcal O}$ is the spectral density of the noisy parameter coupled to $\mathcal O$. For the Floquet qubit, we must modify the formula to account for the time-dependence of the quasi-eigenstates as well as the ambiguity in the quasi-energies $\mathrm{mod}\,\Omega$. Instead we use
\begin{align}\label{eq:floquetbitflips}
    \Gamma_{1\mathcal O} = \frac{1}{\hbar^2}\sum_{k = -\infty}^\infty S_\mathcal{O}(\epsilon_{10} + k\Omega) 
    \left|\overline{\braket{0|\mathcal O e^{ik\Omega t} | 1}}\right|^2
\end{align}
where 
$\overline {(\dots)}$ denotes averaging over one period of the drive \cite{Huang2021}. 

We define the total depolarization rate $\Gamma_1$ from capacitive and inductive loss by summing the individual $\Gamma_{1\mathcal O}$ over all $\mathcal O \in \{\varphi_L, \varphi_R, n_L, n_R\}$, with spectral densities
\begin{align}
    S_{\varphi_j}(\omega) = \frac{2\hbar}{L Q_{ind}(\omega)}\frac{\coth\frac{\hbar|\omega|}{2k_B T}}{1 + \exp \frac{-\hbar \omega}{k_B T}}
\end{align}
for a noisy inductance $L$ and
\begin{align}
    S_{n_j}(\omega) = \frac{2\hbar}{C Q_{cap}(\omega)}\frac{\coth\frac{\hbar|\omega|}{2k_B T}}{1 + \exp \frac{-\hbar \omega}{k_B T}}
\end{align}
for a noisy capacitance $C$, at $T = 15$mK, with assumed frequency-dependent quality factors of 
\begin{align}
    Q_{ind}(\omega) = \left(500 \times 10^6\right) \frac{K_0\left(\frac{h \times 0.5 \textrm { GHz}}{2k_B T}\right)\sinh\left(\frac{h \times 0.5 \textrm { GHz}}{2k_B T}\right)}{K_0\left(\frac{\hbar |\omega| }{2k_B T}\right)\sinh\left(\frac{\hbar |\omega| }{2k_B T}\right)}
\end{align}
for the inductor and
\begin{align}
    Q_{cap}(\omega) = 10^6 \times \left(\frac{6\textrm{ GHz}}{|\omega|/2\pi}\right) ^ {0.7}
\end{align}
for the capacitor \cite{Pop2014,Smith2020}.

By construction, the wavefunctions of the computational states $\psi_0 = \braket{\bm \varphi | 0}$ and $\psi_1 = \braket{\bm \varphi | 1}$ have nearly disjoint support at all times $t$. Thus their matrix element with respect to the error operators $\varphi_{C,D}$ are small, which reduces the rate of depolarization errors. By the same reasoning, transitions from the charge operator $n_j \propto \tfrac{\partial}{\partial \varphi_j}$ are also suppressed.

\subsection{Erasure errors}
Erasure errors, where a computational state decays to an erasure state, follow the same eq. (\ref{eq:floquetbitflips}) as bit-flip errors, but involve different final quasi-eigenstates and thus different matrix elements. The erasure rate due to flux and charge noise, $\Gamma_e$, is roughly constant at $\Gamma_e \approx 1.91$~kHz across the $A,\Omega$ parameter ranges that we contemplate in Figures \ref{fig:perturbationtheory}, \ref{fig:coherence-plot}, \ref{fig:amplitudedephasing}, and \ref{fig:depolcoherence}. These errors are much more likely to occur than bit-flips in the FFM qubit because the relevant matrix elements $\braket{i|\varphi_j|E_k}$ between computational and leakage states are not engineered to be small like the bit-flip matrix elements. We predict that they are by far the dominant source of errors in the qubit, occurring at roughly the kHz level.

Of further interest to error correction is the idea of \textit{state-biased erasures}, where erasure errors are polarized in such a way that the post-erasure state reveals information about the logical pre-error state. This happens when, for instance, $\ket{0}$ preferentially leaks to $\ket{E_0}$ and $\ket{1}$ preferentially leaks to $\ket{E_1}$, or vice-versa. State-biased erasures are even more favorable for error correction than ``normal'' erasures \cite{Sahay2023}, so any state-bias in the FFM erasure rate would further improve its suitability for QEC.

We quantify the state-bias with the parameter $\beta_e$:
\begin{align}
    \beta_e = \frac{\Gamma_e(0\leftrightarrow E_0) + \Gamma_e(1\leftrightarrow E_1)}{\Gamma_e}
\end{align}
where $\Gamma_e(i\leftrightarrow j)$ contains the transition rate between states $\ket{i}$ and $\ket{j}$ only. The cases of $\beta_e = 0$ or $\beta_e = 1$ corresponds to a maximal state-bias, where the post-erasure state completely determines the pre-erasure state, and $\beta_e = \tfrac{1}{2}$ would possess exactly no bias. For the FFM circuit and noise parameters considered so far, $\beta_e \approx 0.45$, indicating a negligible state-bias in our system.

\subsection{Shot noise}\label{shotnoise}
In order to perform dispersive readout operations on the qubit it must be coupled to an external resonator.
Shot noise stems from uncertainty in the photon population in this readout resonator, which dephases the qubit due to the dispersive shift $\chi$. We discuss the dispersive shift in the context of qubit readout in Section \ref{dispersivereadout}, but here it affects the overall phase coherence through the shot noise rate
\begin{align}
    \Gamma_{\phi,\chi} = \frac{ \overline n \kappa}{1 + \kappa^2/\chi^2}
\end{align}
In this formula, $\overline{n} \ll 1$ is the expectation of the resonator photon number and $\kappa$ is the lifetime of the readout resonator \cite{Gambetta2006}. The dispersive shift of our proposed device is roughly comparable to that of the (static) fluxonium molecule studied in \cite{Kou2017}, so we expect the shot noise estimates for the two devices to be comparable.

For our circuit parameters coupled to an 8 GHz readout resonator, we compute $\chi = 0.65$ MHz (see section \ref{dispersivereadout}). Using  $\overline{n} = 10^{-4}$ and $\kappa = 6$ MHz, we find $\Gamma_{\phi,\chi} = 6.95 $ Hz, which is comparable to the predicted dephasing from flux noise in the neighborhood of the double sweet spot.

\section{Single-qubit gates}

We have shown that the FFM device has some ability to protect stored quantum information from decoherence. However, for practical use as a qubit, the device must be able to implement control and readout operations. In this section we show that single-qubit gates can be achieved to high fidelity with external polychromatic flux pulses. 

\begin{figure}
    \centering
    \includegraphics{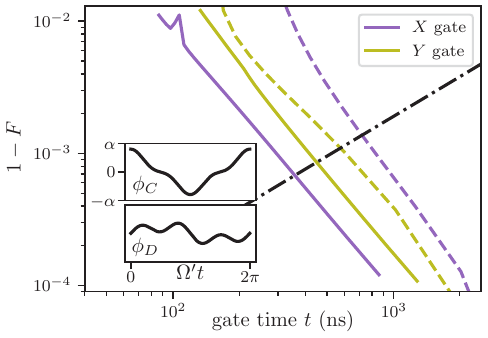}
    \caption{Gate infidelity as a function of gate time for $X = R_{X}(\pi)$ and $Y = R_{Y}(\pi)$ gates induced by a secondary flux drive. The infidelity is minimized (solid lines) by optimizing the mix of secondary drive frequencies, achieving up to 4x faster gates at fixed $1 - F$ compared to a simple unoptimized gate drive which is monochromatic at frequency $\Omega'$ (dashed lines). For both $j = X$ and $Y$ gates, the primary gate axis coefficient satisfies $|c_j|^2 > F - 10^{-7}$. The dash-dotted black line indicates the probability of an erasure error occurring in any interval of length $t$ given the erasure rate from Table \ref{tab:coherence_table} and thus shows a control-independent lower-bound for uncorrected average gate fidelity. The inset shows optimized flux waveforms for the secondary gate flux drive of $\phi_C(\Omega' t)$ and $\phi_D(\Omega' t)$ (relative to the idle values $\phi_C = \pi$ and $\phi_D = 2\pi A \cos\Omega' t$) which implement an $X$ gate with $F = 0.999$. Both insets are on the same axis scale. For this gate, the maximum drive amplitude is approximately $\alpha = A^* / 744$ and the frequency offset is $\delta\Omega/2\pi = -137$ kHz.
    }
    \label{fig:gatefidelity}
\end{figure}

Single-qubit gates can be implemented with an additional drive of the external fluxes $\phi_{C,D}$ beyond the monochromatic drive of $\phi_D$ used to generate the Floquet physics of the qubit. During the steady-state operation discussed thus far, the common flux is fixed at $\phi_C = \pi$ and $\phi_D$ is driven at frequency $\Omega$. To perform an $X$ or $Y$ gate, we must introduce another Hamiltonian term which mixes the computational states without coupling them too strongly to the erasure states or any other excited states of the circuit. 

A gate drive using common mode flux, at the Floquet drive frequency $\Omega$, has a nonzero matrix element between the computational states and is thus used as a starting point for $R_X$ and $R_Y$ gates, from which one can generate all single-qubit gates. In general, a weaker gate drive results in a slower but higher-fidelity rotation gate. Building on these simple monochromatic gates, we show that a polychromatic drive of both $\phi_C$ and $\phi_D$ can further improve gate fidelities and improve gate times. Both types of gates are shown in Figure \ref{fig:gatefidelity}, where we find that optimized polychromatic gates require about $\approx 500$~ns to reach 99.9\% fidelity $X$ and $Y$ gates. On the other hand, simple monochromatic gates require times of $\gtrsim 1$ $\mu$s to reach $\geq 99.9$ \% fidelity for $X$ and $Y$ gates.

Any gate Hamiltonian that approximately generates $X$- rotations in the computational basis has two eigenstates $\ket{\widetilde +}$ and $\ket{\widetilde -}$ that are approximate uniform superpositions of the computational states, with energy difference $\omega_g$. Evolution for time $t$ under this eigensystem produces the approximate rotation gate $R_{\bm n}(\omega_g t)$ around a Bloch sphere axis $\bm n$ determined by $\ket{\widetilde +}$ and $\ket{\widetilde -}$. To bound the gate fidelity for all $t$ in this rotation family, we define the basis change matrix $M_{ij}$ by $M_{ij} = \braket{i | \widetilde j}$ with $i \in \{0,1\}$ and $j \in \{+,-\}$. $M$ is approximately unitary, but fails to be exactly unitary due to residual nonzero support of the gate eigenbasis on leakage and excited states. Under these conditions, we define the fidelity $F$ with respect to the computational basis:
\begin{align}
    F = \sum_{j \in \{x,y,z\}} |\tr M \sigma_z M^{\dagger} \sigma_j/2|^2 = \sum_{j \in \{x,y,z\}} |c_j|^2
\end{align}
This defines the gate axis coefficients $c_{x,y,z}$. Note that the closest unitary gate being applied in this case corresponds to 
rotation around 
$\bm n = (c_x, c_y, c_z)^\top$. The gate time required for an $X$ gate is then $t = \pi/\omega_g$. 

For $Y$ rotations, the situation is exactly analogous, and the eigenstates $\ket{\widetilde{+y}}$ and $\ket{\widetilde{-y}}$ contain the required additional phase.

\subsection{Simple monochromatic gates}
We first consider the fidelity of the simplest possible gate drive protocol, where $\phi_C$ is driven at an amplitude $A_\textrm{gate}$ at frequency $\Omega$. The phase of this drive relative to the Floquet drive of $\phi_D$ sets the gate axis: if the two drives are in-phase, the gate is $R_X$, and if the drives are $\tfrac{\pi}{2}$ out of phase the gate is $R_Y$. A larger $A_\textrm{gate}$ results in a faster gate but with lower fidelity; we show the fidelity vs. gate time for these simple gates in Figure \ref{fig:gatefidelity} in dashed lines.

A bare drive of this form does not result in the highest-fidelity possible gate, though, because the matrix elements of $\varphi_C\cos\Omega t$ (and $\varphi_C\sin\Omega t)$ with excited and erasure states are nonzero and hence result in leakage outside of the computational subspace. The influence of those matrix elements can be suppressed with a more complicated gate drive, as we show next, which can reduced the gate time at fixed fidelity by a factor of $2 - 3$.

However, the simple gates may be practically desirable as they avoid the complications involved in adding additional drive tones. The only parameter is $A_\textrm{gate}$, and small variations in $A_\textrm{gate}$ correspond to small variations in the gate time via Figure \ref{fig:gatefidelity}.

\subsection{Optimized polychromatic gates}

We can improve the situation by introducing more drive tones of both $\phi_C$ and $\phi_D$ at higher harmonics of $\Omega$ and at $\pi/2$ phase offsets. This includes shifts of the drive amplitude $A$, and we also allow shifts of the base frequency $\Omega$. These additional tones can be made to cancel out much of the leakage from the computational subspace, increasing the gate fidelity at a fixed gate time. This problem is relatively well-studied, as multi-tone and multi-operator gates of this form have been studied for quantum optimal control \cite{Werschnik2007} and  DRAG \cite{Theis2018}. In the context of superconducting qubits, quantum optimal control has shown particular promise when the system has disjoint computational wavefunctions, as gates between such states invariably couple to high levels hence require a fine-tuned way to cancel out leakage \cite{Abdelhafez2020,Werninghaus2021}.

We determine the optimal mix of drive tones and frequency offset through optimization. We assume a gate Hamiltonian of the form
\begin{align}
    H_{gate} = H_{FFM}(A^*, \Omega') + \sum_{k=0}^{m_{g}}\,\sum_{j,\theta} x_{jk \theta} \,\varphi_j \cos\left(k\Omega' t + \theta  \right)
\end{align}
where $j \in \{C,D\}$, the phase is $\theta = 0$ or $\pi/2$, and $\Omega' = \Omega^* + \delta\Omega$. This Hamiltonian corresponds to time-dependent external fluxes with Fourier decompositions given by the $x_{jk\theta}$. To find the optimal $X$ gate parameters, for a range of initial amplitudes $A_{gate}$ we fix $x_{C10} = A_{gate}$ and then optimize over the remaining parameters $x_{jk\theta}$ and $\delta \Omega$ to maximize the axis coefficient $|c_x|$. For $Y$ the process is identical except that $x_{C1\tfrac{\pi}{2}} = A_{gate}$ is fixed instead and we maximize $|c_y|$. In practice, optimizing with a fixed $A_{gate}$ corresponds to different gate times even after the optimization is complete. For simplicity we only consider the few lowest Fourier modes by choosing $m_g = 3$; we find that the fidelity is not significantly improved by increasing up to $m_g = 5$. 

The optimization results are shown in Figure \ref{fig:gatefidelity} along with $\phi_C(\Omega't)$ and $\phi_D(\Omega' t)$ waveforms for an $F = 0.999$ $X$-type gate.

\section{Erasure detection and readout}
For use as an erasure qubit, erasure errors must be able to be flagged without disturbing the computational state. In particular, there must be a measurement whose outcome is degenerate with respect to $\ket{0}$ and $\ket{1}$ but whose value is shifted when the FFM is in state $\ket{E_0}$ or $\ket{E_1}$. As we show below, such a measurement is naturally achievable at the double sweet spot operating point by leveraging the second-order insensitivity of the computational quasi-energies --- and simultaneous \textit{sensitivity} of the erasure quasi-energies --- to flux noise.

\subsection{Ancillary qubit erasure detection}\label{ancillaqubitsubsection}
The simplest model for erasure detection is to longitudinally couple the 4-level FFM to an ancillary qubit $q$ via the Hamiltonian
\begin{align}\label{eq:qFmodel}
    H_{qF} = H_{FFM} + \frac{\omega_q}{2} \sigma_z + \frac{g}{2} (\lambda + (1 - \lambda) \sigma_z) \varphi_C,
\end{align}
and use the ancilla qubit frequency as the erasure observable. The interaction term dresses the ancilla-FFM quasi-energies at second order in the coupling strength $g$, and the ancilla frequency is shifted to $\omega_q +  \delta \omega_{q;\beta}$ depending on the FFM state $\beta$. The Pauli operators in these expressions act on the qubit Hilbert space and $\lambda$ is a dimensionless parameter. When the FFM is in the state $\ket{\beta}$, we have the following perturbative expansion:
\begin{align}
    \delta \omega_{q;\beta} &= \delta E(\ket{1_q, \beta_\textrm{FFM}}) - \delta E(\ket{0_q, \beta_\textrm{FFM}}) \\
    &= \frac{g^2}{4}\Bigg(\sum_{\alpha \neq \beta}\frac{|\varphi_C|^2_{\beta\alpha}}{\epsilon_\beta - \epsilon_\alpha} \nonumber \\
    &- (2\lambda - 1)^2 \sum_{\alpha \neq \beta}\frac{|\varphi_C|^2_{\beta\alpha}}{\epsilon_\beta - \epsilon_\alpha}
   \Bigg) + O(g^3)\\
   &= \lambda(1 - \lambda) g^2\sum_{\alpha \neq \beta}\frac{|\varphi_C|^2_{\beta\alpha}}{\epsilon_\beta - \epsilon_\alpha} + O(g^3)
\end{align}

Now let us compare the shift $\delta \omega_{q;\beta}$ between the two cases where the FFM is in a computational state, with $\beta = 1$ or $\beta = 0$. Comparing to equation (\ref{eq:secondordergeneral}) we see that
\begin{align}
    \delta \omega_{q;1}  - \delta \omega_{q;0} = \lambda(1 - \lambda) g^2\xi \Delta^{(2)}_{C} + O(g^3) 
\end{align}
where $\xi$ is a proportionality constant. At the operating point, $\Delta^{(2)}_{C} = 0$ and the ancilla qubit frequency is identical (to order $g^2$) between the two computational FFM states. A similar calculation for the alternate choice of coupling to $\varphi_D$ shows an identical dependence on $\Delta^{(2)}_{D}$, which also vanishes at the operating point. To emphasize that the shift is constant in the logical subspace, we denote $\delta \omega_{q;L} = \delta \omega_{q;0} = \delta \omega_{q;1}$. On the other hand, if $\lambda \neq 0$ and $\lambda \neq 1$ we have in general that $\delta \omega_{q;E_0}  \neq \delta \omega_{q;E_1} \neq \delta \omega_{q;L}$. 

The upshot is that standard spectroscopy on the ancilla qubit \cite{Blais2021} allows one to measure $\delta \omega_{q;\beta}$ and infer whether $\beta = E_0$, $E_1$, or $L$. The measurement thereby reveals whether the FFM is in the logical subspace or in one of the two erasure states, all without potentially disturbing its logical state. Note that a nontrivial shift requires both $\lambda \neq 0$ and $\lambda \neq 1$, but there is no fine-tuning required of the coupling parameters $g$ and $\lambda$: the only fine-tuning necessary to enable erasure detection is that of tuning the drive to the $\Delta^{(2)}_{C} = 0$ line, or at least nearby. At fixed $g$, the shift is maximized if $\lambda = \tfrac{1}{2}$.

Another important property which follows from the above analysis is that, at the double sweet spot operating point, the ancilla may be coupled to any linear combination of $\varphi_C$ and $\varphi_D$ while maintaining the erasure-detection property. This is because both $\Delta^{(2)}_{C}$, $\Delta^{(2)}_{D}$, and the mixed-derivative second order susceptibilities all vanish.

\subsection{Ancillary qubit logical readout}\label{dispersivereadout}
A different coupling to the ancillary qubit used for erasure readout can also enable state discrimination within the logical subspace, enabling readout of the FFM qubit. Instead of a longitudinal coupling, in this case a transverse coupling to the ancillary qubit is required:

\begin{align}\label{eq:qFprimemodel}
    H'_{qF} = H_{FFM} + \frac{\omega_q}{2} \sigma_z + \frac{g}{2} \sigma_x \varphi_C
\end{align}

The dispersive shifts of the ancilla $\delta \omega_{q;\beta}$ are computed in exactly the same way as in section \ref{ancillaqubitsubsection}, but in this case there is no special cancellation that leaves the logical states degenerate. This is because the coupling operator $\sigma_x \varphi_C$ is now off-diagonal on ancilla qubit states, which introduces cross terms in the shift calculation that are generically nonzero. In general, all of the shifts $\delta\omega_{q;\beta}$ for each FFM state $\beta$ are distinct from each other, and in particular the FFM logical states are distinguished from each other during spectroscopic measurement of the ancilla. This enables standard dispersive quantum-non-demolition readout of the FFM qubit \cite{Blais2021}.

\subsection{Ancillary fluxonium model}\label{ancillafluxoniumsubsection}
Remarkably, both readout models in equations (\ref{eq:qFmodel}) and (\ref{eq:qFprimemodel}) can be conveniently implemented with a weak coupling to a single additional fluxonium circuit. Tuning the external flux of this third fluxonium allows one to tune between the longitudinal and transverse couplings required for the different situations of erasure detection and readout.

Concretely, we consider an inductive coupling to an external fluxonium circuit, with Hamiltonian $H_q$, operated in the heavy regime. Denoting the auxiliary flux operator as $\varphi_q$, we consider the inductive coupled total Hamiltonian
\begin{align}\label{eq:coupledancillafluxonium}
    H_{qF} = H_{FFM} + H_q + g_{q}\varphi_q\varphi_C
\end{align}
which can be obtained with a circuit like that shown in Figure \ref{fig:ancillacircuit}.

Two cases are possible depending on the flux detuning $\phi_q$ of the ancillary fluxonium. If $\phi_q$ is detuned significantly from the $0$-flux point, then $\varphi_q \propto \sigma_z + \lambda$ on the lowest two ancilla levels. In this case the coupling is effectively longitudinal, enabling erasure detection. Additionally, $\lambda \approx \tfrac{1}{2}$ in this flux regime, maximizing the dispersive shifts at fixed hybridization of the FFM qubit states (which depends primarily on $g$). This is optimal for maintaining the FFM qubit coherence, as mixing of its states can jeopardize the bit-flip protection; we discuss this shortly in section \ref{ancillacoherence}.

If $\phi_q = 0$, we have $\varphi_q \propto \sigma_x$ and the coupling is transverse, enabling dispersive readout of the logical states. Below, we characterize both situations with exact diagonalization, choosing auxiliary parameters $E_J = 5.2$, $E_C = 0.4$, $E_L = 0.2$~GHz, and coupling $g_q = 0.4$~MHz.

We note that a relatively weak coupling of $g_q = 0.4$~MHz can still give rise to a comparably-sized dispersive shift, even at second order in $g_q$, because (a) the matrix elements of the $\varphi$ operators are generically of magnitude $\sim \pi$ and (b) the energy splittings of the FFM logical and erasure states are of magnitude $\sim 10$~MHz. We can very roughly ballpark the magnitude of the dispersive shifts as
\begin{align}
    |\delta \omega_{q;\beta}| \sim g_q\frac{g_q}{10\textrm{ MHz}} |\pi^2|^2 \sim 4 g_q
\end{align}
and we will see shortly that this is close to the true value.

\subsubsection{Fluxonium erasure detection}
We first set $\phi_q = 0.1\pi$ to investigate the erasure detection properties. These parameters approximately correspond to $\lambda = 0.47$ and $g  = -4.55$~MHz in the effective model of equation (\ref{eq:qFmodel}). Under these circumstances, the ancilla frequency shifts for the logical states are 
\begin{align}
   \delta\omega_{q;0} \approx \delta\omega_{q;1} \approx -2\textrm{ kHz} 
\end{align}
and for the erasure states are
\begin{align}
   \delta\omega_{q;E_0} = -1.99\textrm{ MHz}\quad\quad \delta\omega_{q;E_1} = 1.99\textrm{ MHz}
\end{align}

Moreover, the magnitude of the logical splitting is tiny compared to the erasure splittings: 
\begin{align}
\left|\frac{\delta\omega_{q;0} - \delta\omega_{q;1}}{\delta\omega_{q;E_0}}\right| \approx\left| \frac{\delta\omega_{q;0} - \delta\omega_{q;1}}{\delta\omega_{q;E_1}}\right|\approx 3 \times 10^{-5}
\end{align}

At this external flux, the ancillary qubit has a bare frequency $\omega_q = 3.38$~GHz and a predicted $T_{2,q}$ coherence time of approximately $4.8$~$\mu$s \cite{Groszkowski2018,Chitta2022}. Since $|\delta\omega_{q;E_0}| \approx |\delta\omega_{q;E_1}| \gg \frac{1}{T_{2,q}}$, the erasure shift should be readily detectable. We discuss further coherence impacts of the ancillary qubit in section \ref{ancillacoherence}.

\subsubsection{Fluxonium logical readout}
We now set $\phi_q = 0$ to investigate the logical readout properties, corresponding to $g = 518$~kHz in the ancilla-qubit model of equation (\ref{eq:qFprimemodel}). In this regime, we have ancilla frequency shifts for the logical states of 
\begin{align}
   |\delta\omega_{q;0}|<1\textrm{ kHz} \quad\quad \delta\omega_{q;1} &= -414\textrm{ kHz} 
\end{align}
and for the erasure states of
\begin{align}
   \delta\omega_{q;E_0} &= -25\textrm{ kHz}\quad\quad \delta\omega_{q;E_1} = -174\textrm{ kHz}
\end{align}

At this external flux, the ancillary qubit has a bare frequency $\omega_q = 3.72$~GHz and a predicted $T_{2,q}$ coherence time of approximately $100$~$\mu$s \cite{Groszkowski2018,Chitta2022}, more than large enough to measure the logical shift.

\subsection{Coherence impact of the ancilla}\label{ancillacoherence}
To model the effect of the ancilla on coherence of the FFM, we use the ancilla-qubit model of equation (\ref{eq:qFmodel}). We expect that during normal operation --- when the FFM qubit is not being actively read out --- the attached fluxonium will be left in the longitudinally-coupled regime at $\phi_q \approx 0.1\pi$. Two impacts to FFM qubit coherence are expected: the impact of the new noise experienced by the ancilla, and the impact of the perturbative change to the FFM eigenstates to the already existing FFM noise channels. We show below that the additional ancilla error channels have a minimal effect on bit-flips and dephasing of the FFM qubit, and only a slight impact to the $T_1$ of the FFM qubit due to the perturbative change: we predict that the $T_1$ of the ancilla-FFM coupled system is lower by $1.8\%$ compared to the uncoupled $T_1$ in Table \ref{tab:coherence_table}.

First, we use this model to analyze the impact on the FFM qubit caused by bit-flip and phase-flip noise experienced by the ancilla. Bit-flip noise is expected to be minimal; the expected $T_1$ of the ancilla is above $1$~ms \cite{Zhang2021}, a consequence of its eigenstates being nearly disjoint with respect to its noise operators $\varphi_q$ and $n_q$. We can thus assume that the ancilla remains in its ground state $\ket{0_q}$ between erasure check measurements, which must occur much more frequently than the characteristic FFM erasure time of $T_e \sim 500$~$\mu$s.

In the flux-detuned regime, the ancilla is linearly sensitive to $\phi_q$ noise and so will experience significant dephasing. Conditioned on the qubit being in state $\ket{0_q}$, dephasing manifests as classical noise in the parameter $\omega_q$. However, this noise is not a problem for the FFM qubit because there is no direct coupling in the Hamiltonian between FFM operators and $\omega_q$:
\begin{align}\label{eq:ancilladephasingnoise}
    \braket{0_q | H_{qF} | 0_q} = H_{FFM} - \frac{g}{2}\varphi_C - \frac{\omega_q}{2}
\end{align}

Moreover, the dressed eigenstates of the FFM-ancilla system do not involve hybridization between $\ket{0_q}$ and $\ket{1_q}$ at any order of $g$ in perturbation theory, as the interaction term is diagonal in the $\{\ket{0_q},\ket{1_q}\}$ basis. Therefore, shifts in the relative phase of the states $\ket{0_q}$ and $\ket{1_q}$ do not impact any logical state of the FFM and can only give the FFM state an overall phase. At the level of the ancilla-qubit model, then, there is exactly no dephasing of the FFM due to auxiliary flux noise.

We verify that this holds true in the ancilla-fluxonium model with a numerical finite-difference calculation. Even a large flux excursion of $\delta\phi_q = 2\pi \times 10^{-3}$ causes a frequency shift of the FFM qubit of only approximately $1$~Hz, implying a minimal impact on FFM dephasing even in the more complicated system.

On the other hand, the ancilla interaction causes weak hybridization between the $\ket{1}$ FFM logical state and the erasure states, which do weakly change the coherence properties computed in section \ref{coherence}. Repeating the finite-difference calculations from that section shows no significant change for the dephasing rate or erasure rates, but there is a slight effect on the bit-flip rate, which is raised by a factor of approximately $1.7\%$.

It could be desirable for overall coherence if the ancilla-qubit interaction is activated only when erasure flagging is needed and set to zero otherwise, i.e. if the parameter $g_q$ in equation (\ref{eq:coupledancillafluxonium}) could be tuned dynamically. While this may be possible using a tunable inductive coupler between the FFM and the auxiliary fluxonium \cite{Zhang2024}, the added complication may not be worth the expected marginal coherence benefits, and as such we leave the details of such a scheme to future work.

\section{Outlook}

\begin{figure}
    \centering
    \includegraphics{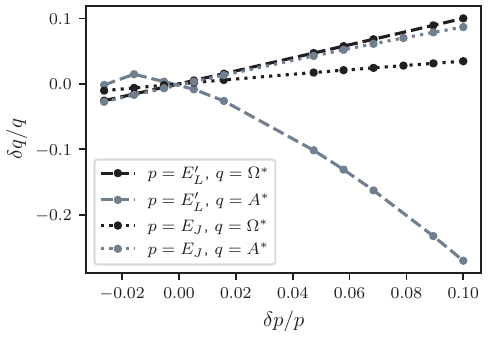}
    \caption{The double sweet spot exists for a wide range of circuit parameters. Here we show the fractional change in the sweet spot parameters $q$ (where  $q \in \{\Omega^*, A^*\}$) as a circuit parameter $p$ is adjusted from its value in Table \ref{tab:circuitparam_table}. We use $p \in \{E_L', E_J\}$ and cutoff values of $N = 50$, $M = 39$, and we computed the location of the double sweet spot through numerical optimization of the coherence time. The flux-noise dephasing rate remains low at each point, satisfying $\Gamma_{\phi, \varphi_D} + \Gamma_{\phi, \varphi_C} \leq 1$~Hz.}
    \label{fig:paramsweep}
\end{figure}

In summary, we have presented a novel type of superconducting qubit, the FFM qubit, which uses a strong flux drive to suppress dephasing while preserving a high $T_1$. The procedure unavoidably introduces two additional erasure states.  For the circuit parameters that we have explored, the dephasing error rate is an order of magnitude less then the depolarizing error rate, which in turn is two orders of magnitude less then the erasure error rate. It might be possible to decrease both the depolarization and erasure error rates somewhat by further tuning circuit parameters. Additionally, the bias toward erasure errors is favorable for quantum error correction \cite{Kubica2023}.

In addition to the coherence analysis, we have optimized waveforms for flux-driven single qubit rotation gates. We leave the implementation of multi-qubit gates to future work.

The existence of a double sweet spot is not unique to the circuit parameters in Table \ref{tab:circuitparam_table}; indeed, the four-level coherence analysis in section \ref{4leveltheory} relies only on the assumption that $E_L, E_C \ll E_J$. We emphasize this point in Figure $\ref{fig:paramsweep}$, which shows how the sweet spot parameters $A^*, \Omega^*$ move as the two circuit parameters $E_J$ and $E_L'$ are independently tuned.

Lastly, we point out that the design motivation may be more broadly applicable to other types of qubit hardware. The key ingredient to achieving the low error rates is enhanced tunneling between the two non-computational states of two coupled qubits that exhibit localized, disjoint eigenstates. This tunneling is obtained by selectively coupling them with a resonant drive, delocalizing them and lifting degeneracy with the still-disjoint computational states; the result is a non-divergent and even potentially vanishing $\Delta^{(2)}$. Although we have discussed a concrete implementation with fluxonium and circuit QED, this general formula could be applied to different physical systems of coupled qubits. Spin qubits in particular could be an attractive platform as their Hilbert space is finite-dimensional, avoiding the problem of multi-photon resonances from strong drives.

\begin{acknowledgments}
We acknowledge support from the NSF Quantum Leap Challenge Institute for Hybrid Quantum Architectures and Networks (NSF Award 2016136) (B.K.C. and M.T.). This work made use of the Illinois Campus Cluster, a computing resource that is operated by the Illinois Campus Cluster Program (ICCP) in conjunction with the National Center for Supercomputing Applications (NCSA) and which is supported by funds from the University of Illinois at Urbana-Champaign. This research was supported in part by the National Science Foundation under Grant No. NSF PHY-1748958 and by the Heising-Simons Foundation (M.T.). This work is partially supported by the Air Force Office of Scientific Research under award number FA9550-21-1-0327 (A.K.).
\end{acknowledgments}

\bibliography{main}

\onecolumngrid
\appendix

\section{The Floquet frequency lattice}\label{floquetlattice}
In the following sections we will explain how to diagonalize the Floquet Hamiltonian $K_4$ of the FFM qubit considering the lowest four levels of the static Hamiltonian $H_{dc}$ only. In this case, $H_{dc}$ is equal to $H_{FFM}(t = 0)$ with $A = 0$, or equivalently eq. (\ref{eq:FFMHam}) with constant $\phi_L(t) = \phi_R(t) = \pi$.

The Floquet theorem \cite{ASENS_1883_2_12__47_0} guarantees the existence of a set of states $\ket{\psi_{\alpha}(t)}$ of the following form that, at all times $t$, both form a complete basis of states and solve the Schrodinger equation:
\begin{align}
    \ket{\psi_{\alpha}(t)} &= e^{-i \epsilon_\alpha t}\ket{q_\alpha(t)} \\
    i\partial_t \ket{\psi_{\alpha}(t)} &=  H_{FFM}(t)\ket{\psi_{\alpha}(t)}
\end{align}
such that the states $\ket{q_\alpha(t)}$ are periodic with period $\tau = 2\pi/\Omega$. These states are referred to as quasi-eigenstates and the $\epsilon_\alpha$ are known as quasi-energies; a consequence of the definition is that the $\epsilon_{\alpha}$ are only uniquely defined $\mathrm{mod}\,\Omega$. 

An equivalent formulation, known as the frequency lattice, defines a new Hermitian operator $K$ over a larger Hilbert space and encodes the quasi-eigenstates and -energies in its eigenvectors and eigenvalues. As the $\ket{q_\alpha}$ are periodic, we may Fourier-expand them:
\begin{align}
    \ket{q_\alpha(t)} = \sum_{n = -\infty}^\infty e^{in\Omega t}\ket{\phi_{n\alpha}}
\end{align}
where the Fourier components $\ket{\phi_{n\alpha}}$ are un-normalized. Similarly we denote the $n$-th Fourier component of $H_{FFM}$ as $\widetilde H_{n}$. Then, integrating the Schrodinger equation over one period yields an eigenvalue equation for the $\ket{\phi_{j\alpha}}$:
\begin{align}\label{eq:flo-K}
    K_{ij} &= \widetilde H_{i-j} + j\Omega \delta_{ij} \\ \left(\bm{\phi}_{\alpha}\right)_j &= \ket{\phi_{j\alpha}} \\    \epsilon_{\alpha}\bm{\phi}_{\alpha} &= K \cdot  \bm{\phi}_\alpha
\end{align}
Here, we have defined the Floquet Hamiltonian $K$, which is a block matrix: the blocks are labeled by the Fourier indices $(i,j)$, and within a block the matrix acts on the Hilbert space of $H_{dc}$. We will find the time-dependent quasi-eigenstates by solving for their Fourier expansions, the eigenvectors $\bm\phi_\alpha$ of $K$.

\section{Diagonalizing $K_{FFM}$}\label{diagonalizeK}
We now proceed with analyzing $H_{FFM}$, with $\phi_D(t) = 2\pi A \cos \Omega t$ and $\phi_C = \pi$. By the Floquet theorem, this has identical quasi-energies to the alternative phase choice of $\phi_D \propto \sin \Omega t$, although the quasi-eigenstates will carry different phases. This choice of $H_{FFM}$ has the following Fourier components:
\begin{align}
    \widetilde H_{k} = \begin{cases}
     H_{dc} & k = 0 \\
     \tfrac{A\pi}{4}\Delta E\, \varphi_D & |k| = 1 \\
     \tfrac{A^2\pi^2}{8}\Delta E & |k| = 2 \\
     0 & |k| > 2
    \end{cases}
\end{align}
Here, $A$ is the differential flux drive amplitude and $\Delta E = 2E_L - E'_L$. We will project to the subspace $\mathcal H_4$ spanned by the four lowest eigenstates of $H_{dc}$, labeled $g,e,h,f$ with energies $0,\delta,\Delta, \mu$ respectively. In this basis $B'$ we have the matrices
\begin{align}
(H_{dc})_{B'} &= \left(
\begin{array}{cccc}
  0&  &  &  \\
  & \delta  &  &  \\
  &  & \Delta  &  \\
  &  &  & \mu  \\
\end{array}
\right) \label{eq:bprimebasis} \\
(\varphi_D)_{B'} &= \varphi_0 \left(
\begin{array}{cccc}
 0 & 1 & 0 & 0 \\
 1 & 0 & -\sqrt{2} \epsilon  & 0 \\
 0 & -\sqrt{2} \epsilon  & 0 & 0 \\
 0 & 0 & 0 & 0 \\
\end{array}
\right) \label{eq:phidbprimebasis}
\end{align}
Equation (\ref{eq:phidbprimebasis}) defines the parameters $\varphi_0$ and $\epsilon$, and we will assume that $\epsilon$ is small in order to use it as a perturbative parameter. For the parameters in Table \ref{tab:circuitparam_table} we compute $\epsilon = 0.0437$.

It will be more convenient to work in an eigenbasis of $\varphi_D$ restricted to $\mathcal H_4$. We label these states $x,y,z,w$, and in this basis $B$ we obtain the following matrices:
\begin{align}
    (\varphi_D)_B &= \varphi_0 \left(
\begin{array}{cccc}
  -\sqrt{1 + 2\epsilon^2}&  &  &  \\
  & \sqrt{1 + 2\epsilon^2}  &  &  \\
  &  & 0  &  \\
  &  &  & 0 \\
\end{array}
\right) 
\\\label{eq:HdcBbasis}
(H_{dc})_B &=\Delta\left(
\begin{array}{cccc}
 \epsilon ^2 + \frac r 2 &  \epsilon ^2-\frac r 2 & - 
   \epsilon  & 0 \\
  \epsilon ^2-\frac r 2  & \epsilon ^2  + \frac r 2 & - 
   \epsilon  & 0 \\
 -  \epsilon  & -  \epsilon  & 1-2 \epsilon ^2 & 0 \\
 0 & 0 & 0 & \mu/\Delta  \\
\end{array}
\right)
\end{align}
Here we have put $r = \delta/\Delta$. Notice that the $f$ state was decoupled from the other rows of the $\varphi_D$ matrix in equation (\ref{eq:phidbprimebasis}), so we have taken $\ket{w} = \ket{f}$. 

This is a desirable situation, as we can now do perturbation theory in $\epsilon$ and $r$. All terms of order zero in these parameters are on the diagonal of both $H_{dc}$ and $\varphi_D$, which means that the zeroth order Floquet Hamiltonian $K^{(0)}$ is exactly solvable. In fact it is the direct sum of two infinite 5-diagonal matrices $X$ and $Y$ as well as two infinite tri-diagonal matrices $Z$ and $W$:
\begin{align}
    X_{ij} = -z_0 \Omega(\delta_{i,j-1} + \delta_{i,j+1}) + z_1 \Omega(\delta_{i,j-2} + \delta_{i,j+2}) + j\Omega \delta_{ij} \\
    Y_{ij} = z_0 \Omega(\delta_{i,j-1} + \delta_{i,j+1}) + z_1 \Omega(\delta_{i,j-2} + \delta_{i,j+2}) + j\Omega \delta_{ij} \\
    Z_{ij} =  z_1 \Omega(\delta_{i,j-2} + \delta_{i,j+2}) + (j\Omega + \Delta) \delta_{ij} \\
    W_{ij} =  z_1 \Omega(\delta_{i,j-2} + \delta_{i,j+2}) + (j\Omega + \mu) \delta_{ij}
\end{align}

where the indices $i,j$ range over all integers. We have introduced the normalized drive amplitudes $z_0 = A/A_0$ and $z_1 = (A/A_1)^2$ with
\begin{align}\label{eq:a0eq}
    1/A_0 &=  \frac{\pi\varphi_0\sqrt{1 + 2\epsilon^2}}{2}\frac{\Delta E}{\Omega} \\
    1/A_1^2 &= \frac{\pi^2\varphi_0}{4} \frac{\Delta E}{\Omega}
\end{align}

The diagonalizations of these matrices are known. In each case, the eigenvalues are the diagonal entries: for each integer $n$, they are $n\Omega$ for $X$ and $Y$, $n\Omega + \Delta$ for $Z$, and  $n\Omega + \mu$ for $W$. The eigenvector labeled by the integer $n$, denoted by the corresponding lowercase letter (for instance, $\ket{\widetilde x,n}$ for the $X$ matrix), is, for each of the four matrices,

\begin{align}
    \ket{\widetilde x, n} &= \sum_{k= -\infty}^\infty\sum_{m= -\infty}^\infty J_{m}(z_1) J_{k - n -2m}(-z_0)
    \ket{x,k} \\
    \ket{\widetilde y, n} &= \sum_{k= -\infty}^\infty\sum_{m= -\infty}^\infty J_{m}(z_1) J_{k - n -2m}(z_0)
    \ket{y,k} \\
    \ket{\widetilde z, n} &= \sum_{m = -\infty}^\infty J_{m}(z_1) 
    \ket{z,2m + n} \\
    \ket{\widetilde w, n} &= \sum_{m = -\infty}^\infty J_{m}(z_1) 
    \ket{w,2m + n}
\end{align}
where for instance $\{\ket{x,n}\}_{n\in \mathbf{Z}}$ labels the standard basis of the matrix $X$ and so on for $Y,Z,W$. Since $K^{(0)} = X \oplus Y \oplus Z \oplus W$, these four collections of states together constitute an eigenbasis for $K^{(0)}$.

At this point we will make the assumption that $\Omega \approx \Delta$, which means that for each integer $n$ (at order zero in $\epsilon$ and $r$) there is a triplet of nearly degenerate states: in particular, $S_n = \mathspan\{\ket{\widetilde x, n}, \ket{\widetilde y, n}, \ket{\widetilde z, n-1}\}$ is a nearly degenerate subspace under $K^{(0)}$. Additionally, the $W$ eigenstates have all zero matrix elements with the $r$ and $\epsilon$ perturbations, so we are finished with them for now and will restrict our attention to the $S_n$ subspaces. In particular, we use the Generalized Van Vleck (GVV) formalism to perform nearly-degenerate perturbation theory in each subspace.

As a simplifying assumption, we will work only in one perturbative parameter $\epsilon$ by setting $r = R \epsilon^2$ with $R$ fixed. In principle this is not necessary for the GVV formalism, but the assumption is valid for a wide range of device parameters, including the ones we consider numerically, and can if necessary be lifted on a case-by-case basis. For our parameters $R = 1.148$.

According to the GVV formalism \cite{Son2009}, we form the effective Hamiltonian matrix up to order $\epsilon^2$, along with the basis vectors $\mathbf b_j$ ($j = 1,2,3$) up to order $\epsilon$:

\begin{align}
    G &= \sum_{m} \epsilon^m G^{(m)} \\
    \bm b_j &= \sum_{m} \epsilon^m \bm b_j^{(m)} \\
    \bm b^{(0)} &= \begin{pmatrix}
        \ket{\widetilde x, n} &  \\
        \ket{\widetilde y, n} & \\
        \ket{\widetilde z, n-1} &
    \end{pmatrix}
\end{align}

The relevent higher-order terms are
\begin{align}
    \bm b_j^{(1)} &= R_j K^{(1)} \cdot \bm b_j^{(0)} \\
    G^{(0)} &= \bm b^{(0)} \cdot K^{(0)} \cdot  \bm b^{(0)} \\
    G^{(1)} &= \bm b^{(0)} \cdot K^{(1)} \cdot \bm b^{(0)} \\
    G^{(2)} &= \bm b^{(0)} \cdot K^{(2)} \cdot \bm b^{(0)} + \bm b^{(0)} \cdot K^{(1)} \cdot \bm b^{(1)} - G^{(1)} \cdot \left(\bm b^{(0)} \bm b^{(1)}\right)
\end{align}
where $R_j = \sum_{\phi}\frac{\ket{\phi}\bra{\phi}}{E_j - E_\phi}$ is the resolvent.

However, we will ultimately be interested in a low-amplitude regime. By working to zeroth order in $z_1 \propto A^2/32$, we can simplify the algebra considerably.

\section{Low-amplitude approximation}\label{lowampappendix}

To simplify the analysis we will now make the approximation of $z_1 \ll 1$. Working to zeroth order in $z_1$, the states in $S_n$ are simplified:

\begin{align}
    \ket{\widetilde x, n} &= \sum_{k= -\infty}^\infty J_{k - n }(-z_0)
    \ket{x,k} \label{eq:lowamp_xyzwstates_x} \\ 
    \ket{\widetilde y, n} &= \sum_{k= -\infty}^\infty J_{k - n }(z_0)
    \ket{y,k} \label{eq:lowamp_xyzwstates_y} \\ 
    \ket{\widetilde z, n} &= \ket{z,n} \label{eq:lowamp_xyzwstates_z}\\ 
    \ket{\widetilde w, n} &= \ket{w,n} \label{eq:lowamp_xyzwstates_w}
\end{align}

Using this, the $G^{(k)}$ matrices can be computed for each integer $n$:
\begin{align}
    G^{(0)} &= \left(
\begin{array}{ccc}
 n\Omega  & 0 & 0 \\
 0 & n\Omega  & 0 \\
 0 & 0 & \Delta + (n-1)\Omega  \\
\end{array}
\right) \\
G^{(1)} &= \Delta   J_1(z_0)\left(
\begin{array}{ccc}
 0 & 0 & 1 \\
 0 & 0 & -1 \\
 1 & -1 & 0 \\
\end{array}
\right) \\
G^{(2)} &= \Delta \left(
\begin{array}{ccc}
  \left(\frac{R}{2}+1 \right)+\  f_-  &  \left(1 -\frac{R}{2}\right) J_0(2 z_0)+
  f_+  & 0 \\
  \left(1 -\frac{R}{2}\right) J_0(2 z_0)+ f_+  &  \left(\frac{R}{2}+1\right)+
   f_-  & 0 \\
 0 & 0 &  -2  (1 + f_-)  \\
\end{array}
\right)
\end{align}
The functions $f_+$ and $f_-$ are defined as
\begin{align}\label{eq:fplusminus}
    f_{\pm} = -\sum_{k \neq 0} \frac{J_{1-k}(z_0)J_{k-1}(\pm z_0)}{k}
\end{align}
The eigenvectors of $G^{(\leq 2)} = G^{(0)} + \epsilon G^{(1)} + \epsilon^2 G^{(2)}$ are
\begin{align}
    \bm{e}_0 &= \frac{1}{\sqrt{2}}\begin{pmatrix}
        1 & 1 & 0
    \end{pmatrix}^\top \\
    \bm{e}_1 &= \frac{1}{\mathcal N_1}\begin{pmatrix}
        (\alpha_1 - \alpha_2) & (\alpha_2 - \alpha_1) & 1
    \end{pmatrix}^\top \\
     \bm{e}_2 &= \frac{1}{\mathcal N_2}\begin{pmatrix}
        (\alpha_1 + \alpha_2) & (-\alpha_1 - \alpha_2) & 1
    \end{pmatrix}^\top \\
    \alpha_1 &= \frac{\epsilon \big((R-2) J_0(2 z_0)+R + 2(3 f_- - f_+ +3)\big) + 2 \epsilon^{-1}(\Omega/\Delta - 1) }{8  J_1(z_0)} \\
   \alpha_2 &=\sqrt{\alpha_1^2 +\frac{1}{2}}
\end{align}
Here $\mathcal N_1$ and $\mathcal N_2$ are normalizations fixed by $\|\bm{e}_j\| = 1$. We choose the state corresponding to $\bm{e}_0$ (with $n = 1$) to be the computational state $\ket{0}$, and the $\ket{w,0}$ state to be the computational state $\ket{1}$. The other two states corresponding to $\bm{e}_1$ and $\bm{e}_2$ are then the erasure states.

The $\ket{0}$ computational state is
\begin{align}\label{eq:lowamp_zerostate}
    \ket{0} = \frac{1}{\sqrt{2}}\left(\ket{\widetilde x, 0} + \ket{\widetilde y, 0} \right) + O(\epsilon)
\end{align}

As we now have expressions for the erasure states, we can compute their matrix elements with respect to $\varphi_C$ and $\varphi_D$, and thus calculate the second-order dephasing contribution $\Delta^{(2)}$. We find the following conditions on $\Omega^{*j}$, the drive frequency where $\Delta^{(2)}_{j} = 0$ for $j = \varphi_D$ or $j = \varphi_C$ respectively:
\begin{align}
    \Omega^{*D} &= \Delta - \Delta\epsilon ^2 \left(3 + 3f_- + f_+ + \tfrac R 2 + J_0(2z_0)(1 - \tfrac R 2)) \right) \\
   \Omega^{*C} &=  \mu - \Delta \epsilon^2 \left(1 + f_-
   - f_+ + \tfrac R 2 - J_0(2 z_0)(1 - \tfrac R 2)\right)
\end{align}
These curves in the $(z_0,\Omega)$ plane may cross zero or more times depending on the parameters $R$ and $\Delta - \mu$. If it exists, we denote the point at which they cross with minimal $z_0$ as $(z_0^*, \Omega^*)$ which through equation (\ref{eq:a0eq}) gives $(A^*, \Omega^*)$.

The qubit frequency is given by
\begin{align}
    \epsilon_{10} = \mu - \Omega - \Delta \epsilon^2 \left(1+ f_- +  f_+ + \tfrac R 2 + J_0(2 z_0)(1 - \tfrac R 2)\right)
\end{align}

\begin{figure}
    \centering
    \includegraphics{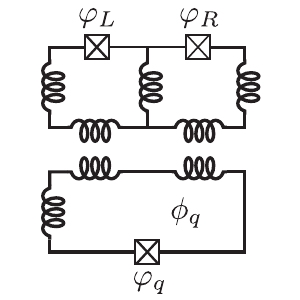}
    \caption{Lumped-element circuit model for the ancillary fluxonium which enables erasure detection and logical readout.}
    \label{fig:ancillacircuit}
\end{figure}

\begin{figure}
    \centering
    \includegraphics{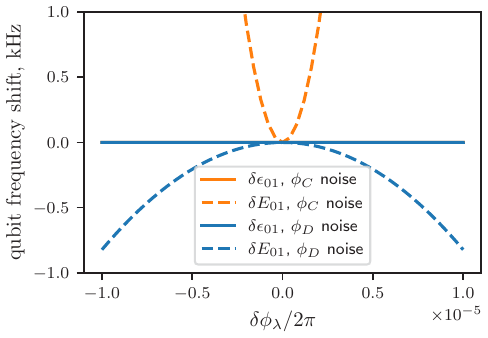}
    \caption{Dispersion of the FFM qubit frequency $\epsilon_{01}$ at the approximate double sweet spot point $(A^*, \Omega^*)$ (solid), and of the FM frequency $E_{01}$ (dashed), from their noise-free values as flux noise is applied. Both common and differential flux noise is considered; the FFM frequency response is extremely flat and cannot be distinguished from zero on this scale for either noise channel. This can be seen as a consequence of the $\Delta^{(2)}_C \approx \Delta^{(2)}_D \approx  0$ property at the double sweet spot for the FFM, whereas those coefficients are large for the FM.}
    \label{fig:dispersion}
\end{figure}

\section{Numerical simulation convergence}\label{convergence}

\begin{figure}
    \centering
    \includegraphics{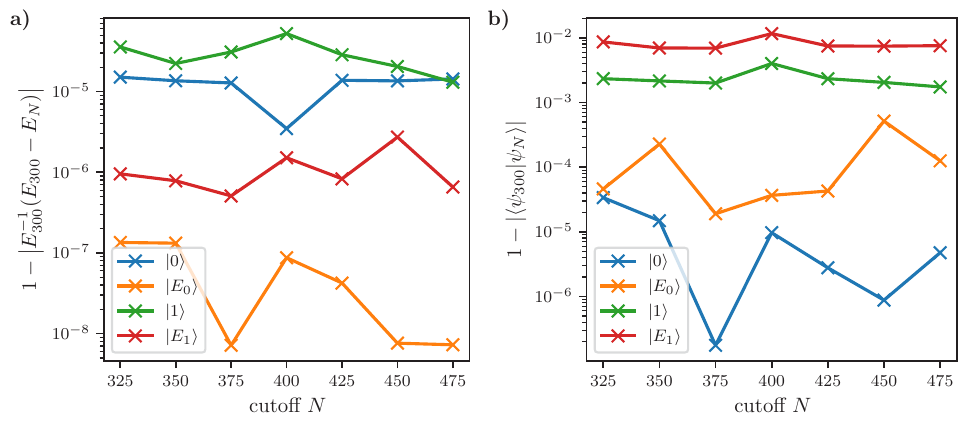}
    \caption{Convergence of our numerical simulations, showing the (a) relative quasi-energy change and (b) change in the overlap of the quasi-eigenstates as the static cutoff $N$ is tuned from its nominal value of $300$. The Fourier cutoff of the quasi-eigenstates (which are exponentially localized in Fourier space) is fixed at $M = 63$.  The largely independent and small differences as a function of the static cutoff indicate that the simulation results at $N = 300$ are relatively well converged.}
    \label{fig:convergence}
\end{figure}

The wavefunctions and coherence data in Figures 1-3, 5, and 6 are the result of numerical diagonalization of the Floquet frequency lattice Hamiltonian $K$ (see Appendix \ref{floquetlattice}) with $M = 63$ Fourier modes (ranging from $-31$ to $31$) and the lowest $N = 300$ eigenstates of the static $H_{FM}$. Figure \ref{fig:gatefidelity} instead used $N = 400$ static eigenstates. In all cases eigenstates of $H_{FM}$ were obtained by diagonalizing  eq. (\ref{eq:FFMHam}) with $\phi_C = \pi$ and $\phi_D = 0$ in the harmonic oscillator basis where $\phi_{L,R} = \frac{x_0}{\sqrt{2}}(a_{L,R} + a_{L,R}^\dagger)$, with $x_0 = \sqrt[4]{8 E_C/E_L}$, using 100 basis states each of $a_L$ and $a_R$ for a total of $10^4$ ladder operator basis states.

We provide evidence that our simulation of the computational and erasure states is converged in Figure \ref{fig:convergence} where we tune the static cutoff $N$ from 300 to 475. The Fourier cutoff is fixed at $M = 63$, which we expect provides a good approximation due to exponential Wannier-Stark localization in the frequency lattice \cite{Wannier1962}.

\end{document}